\documentclass[preprint,eqsecnum,aps,showpacs]{revtex4}
\usepackage{dcolumn}
\usepackage{graphicx}
\usepackage{latexsym}

\begin{document}

\title{Particle creation in the presence of a warped extra dimension} 

\author{Suman Ghosh\footnote{Electronic address : {\em
suman@cts.iitkgp.ernet.in}}${}^{}$ and Sayan Kar\footnote{Electronic
address : {\em sayan@cts.iitkgp.ernet.in}}${}^{}$}
\affiliation{Department of Physics and Centre for Theoretical Studies
\\ Indian Institute of Technology, Kharagpur 721 302, India}

\begin{abstract}
Particle creation in spacetimes with a warped
extra dimension is studied. In particular, we investigate 
the dynamics of a conformally coupled, massless scalar field 
in a five dimensional warped geometry where the induced 
metric on the 3--branes is that of a
spatially flat cosmological model. We look at situations where
the scale of the extra dimension is assumed (i) to be time independent 
or (ii) to have specific functional forms for time dependence. 
The warp factor is chosen to be that of the Randall--Sundrum model. 
With particular choices for the functional form of the 
scale factor (and also the function characterising the time
evolution of the extra dimension) we obtain 
the ${\vert \beta_k\vert}^2$, the particle number and energy
densities after solving (wherever possible, analytically but, otherwise,
numerically) the conformal scalar field equations. 
The behaviour of these quantities for the massless 
and massive Kaluza--Klein modes are examined.
Our results show the effect of a warped extra dimension on
particle creation and illustrate how the nature of particle
production on the brane depends on 
the nature of warping, type of cosmological evolution as well as the
temporal evolution of the extra dimension.    
\end{abstract}


\maketitle


\section{Introduction}

Beginning with Kaluza-Klein \cite{kk}, a large variety of models with
extra dimensions have been proposed over the years. The recent
brane-world models \cite{wbw1}-\cite{wbw4} where our world
is viewed as a four dimensional sub-manifold (3-brane)
embedded in five dimensions are actively pursued today
largely because of their potential in proposing achievable
experimental signatures of extra dimensions.  
Among braneworld models, the warped type necessarily assumes 
a curved higher dimensional spacetime and the line element on the
3--brane is {\em scaled} by a {\em warp factor}, thereby rendering the
higher dimensional metric non-factorisable. 
The brane-world models seem to provide a viable resolution of the 
long-standing hierarchy problem in high energy physics. The warped 
braneworlds also suggest a dynamic way of {\em compactification}
by proposing the idea of localisation of fields on the brane \cite{loc} 
(and references therein).

A study of quantum field theory in the context of warped
spacetimes where there is an extra dimension is therefore not
inappropriate. However, not much has been done along this direction.
Leaving aside warping, the analysis of 
quantum field theory in a higher dimensional
spacetime with a Kaluza--Klein-like extra dimension has been discussed by
some--notable among them being the works reported in 
 \cite{garriga}, \cite{nojiri}, \cite{makharko}, {\cite{huang}.
More recently, Saharian \cite{saharian} (and references therein) has 
discussed quantum fields
in such spacetimes quite extensively (though, somewhat on the formal side)
in a series of papers. Gravitational particle production in braneworld
cosmology and its implications have been studied in \cite{urban}.  

The questions we address in this article are the following.
Particle creation in a cosmological background spacetime using the
formalism of quantum fields in curved spacetime \cite{birrell} is
a well--studied subject. How do the presence of extra dimensions
affect particle creation? Further, how does a warping of the higher dimensional
spacetime create differences, if any. The simple answer to the first question 
is related to the fact that 
 now we do not refer to ${\vert \beta_k \vert }^2$ but we must
consider ${\vert \beta_{k,k_5}\vert}^2$, where $k_5$ is the
momentum associated with the extra dimension. The eigenvalues 
$k_5$ would have to be obtained by solving the equation for the 
extra dimensional
part of the field (assuming separability) and the corresponding 
equation (say, for a scalar or a vector or a fermion). In the
case of a KK like extra dimension $k_5 = \frac{n}{R_5}$ where $R_5$ is
related to the radius of the compact extra space (say a circle). Not so
when we have a warped spacetime. Here, the equation for the 
extra dimensional piece of the field  would be different for 
different types of warping and
hence, the resulting solutions and eigenvalues will, obviously differ.
Further, in a two brane scenario, appropriate boundary conditions need to be
imposed and thus $k_5$ would take on only those allowed 
values, such that the boundary
 conditions are obeyed. The question therefore comes up: how do the values for
$k_5$ as well as the nature of evolution of the scale factor and the extra
dimensions affect particle creation? We shall provide illustrations to this 
end in the rest of this article. In addition, we also consider the
case of a time--dependent extra dimension. In this context, we investigate
how the nature of time evolution of the extra dimension affects
particle creation characteristics.  

The plan of the article is as follows.
In Section II, we discuss the conformally 
coupled scalar field equation and also the numerical method of solving
the equations. The analytic formalism (for four dimensional cosmological 
spacetimes) is well-known and given 
in \cite{birrell}. Section III contains the analysis of the 
extra dimensional part (assuming separability) of the scalar field equation. 
Then, in Section IV, we find out the $\vert \beta_k^2\vert$, the number
and energy density of the created particles for the 
four dimensional scenario with a massive, conformally coupled scalar field.
Sections V and VI deal with the similar analysis for a massless scalar field 
in the presence of a 
time--independent extra dimension and a time dependent extra dimension, 
respectively. Section VII discusses the thermal/non--thermal nature of the 
spectra.
Section VIII analyses zero mode particle creation
and, finally, in Section IX, we conclude with comments and suggestions
on future work.

\section{Quantum field coupled to a spacetime with an extra dimension}  

Let us consider the background line element (using conformal time) to be 
generically of the form:
\begin{equation} 
ds^2 = e^{2f(\sigma)} a^2(\eta)[- d\eta^2 + dx^2 + dy^2 + dz^2] + \phi^2(\eta) d\sigma^2 ,\label{eq:metric}
\end{equation}
where $ e^{2f(\sigma)}$ is the warp factor, $a(\eta)$ and $\phi(\eta)$ are
the scale factors associated with the ordinary space (${\vec x}$) and the extra
dimension ($\sigma$) respectively. $\eta$ denotes conformal time.
 
A scalar field $\psi(\eta,\vec{x},\sigma)$ conformally 
coupled to the above metric
satisfies the following Klein-Gordon equation,
\begin{eqnarray}  
& & -\ \frac{e^{-2f(\sigma)}}{a^2} \frac{\partial^2 \psi}{\partial \eta^2} + \frac{e^{-2f(\sigma)}}{a^2} \left(\frac{\partial^2 \psi}{\partial x^2} + \frac{\partial^2 \psi}{\partial y^2} + \frac{\partial^2 \psi}{\partial z^2}\right) + \frac{1}{\phi^2} \frac{\partial^2 \psi}{\partial \sigma^2}\nonumber \\ & & -\ \frac{e^{-2f(\sigma)}}{a^2} \left (\frac{2\dot a}{a} + \frac{\dot \phi}{\phi} \right )\frac{\partial \psi}{\partial \eta} + \frac{4f'}{\phi^2} \frac{\partial \psi}{\partial \sigma} - (m^2 + \xi R)\psi = 0 ,\label{eq:genel} 
\end{eqnarray}
where $m$ is the mass of the scalar particle, $\xi$ is the conformal coupling
constant and $R$ is the five-dimensional curvature scalar. A dot ($
^.$) denotes differentiation w.r.t $\eta$ and a prime ($\prime$)
denotes differentiation w.r.t $\sigma$.

We now concentrate on conformally coupled ($\xi =
\frac{3}{16}$ in five dimensions) massless particles ($m = 0$). 
The scale factor evolution, the time-dependent extra dimension (we also
discuss the time--independent case later)
and the warp factor lead to distinct characteristics of  particle creation.

We separate variables using the following ansatz for the scalar field:  
\begin{equation}
 \psi (\eta,{\bf x},\sigma) = \frac{1}{{e^{f} a \phi^{\frac{1}{2}} }} \chi_l (\eta)F({\bf x})G(\sigma) .
 \label{eq:separating}
\end{equation}


The normalization condition for $\psi$ gives the Wronskian relation,
\begin{equation}
\dot \chi_l^*\chi_l - \dot\chi_l\chi_l^* = i  .\label{eq:wronskian}
\end{equation}
Let 
\begin{equation}
\frac{1}{F({\bf x})}\left\{ \frac {d^2F({\bf x}) }{dx^2}+ \frac {d^2F({\bf x})}{dy^2} + \frac {d^2F({\bf x})}{dz^2} \right\} = -{\bf k}^2 ,\label{eq:separatingx}
\end{equation}

and,
\begin{equation}
e^{2f} \left\{\frac {G''(\sigma)}{G(\sigma)} + 2 f'\frac{G'(\sigma)}{G(\sigma)} + \left(\frac{f''}{2} + \frac{3f'^2}{4}\right) \right\} = - k_\sigma^2 .
\label{eq:separatingsigma}
\end{equation}

The above two assumptions imply,
%
\begin{equation}
\ddot \chi_l(\eta) + \left[ \left({\bf k}^2 + \frac{a^2}{\phi^2} k_\sigma^2 \right) + \frac{\ddot a}{8a} - \frac{\ddot \phi}{8\phi} + \frac{\dot \phi^2}{4\phi^2} - \frac{\dot a\dot \phi}{4a\phi} \right]\chi_l(\eta) = 0 .
\label{eq:genchieqneta}
\end{equation}


One can write Eq.(\ref{eq:genchieqneta}). as,
\begin{equation}
\ddot \chi_l(\eta) + \left[ \Omega_l^2 + Q \right]\chi_l(\eta) = 0 ,\label{eq:genform}
\end{equation}
where,
\begin{equation}
\Omega_l^2(\eta) = \left({\bf k}^2 + \frac{a^2}{\phi^2} k_\sigma^2 \right)\label{eq:Omgforms}
\end{equation}
and
\begin{equation}
Q(\eta) =\frac{\ddot a}{8a} - \frac{\ddot \phi}{8\phi} + \frac{\dot \phi^2}{4\phi^2} - \frac{\dot a\dot \phi}{4a\phi}.\label{eq:Qforms}
\end{equation}

This equation admits WKB solutions of the form,
\begin{equation}
\chi_l = \frac{\alpha_l}{\sqrt{2\Omega_l}} e^{-i\int^\eta \Omega_l d\eta} + \frac{\beta_l}{\sqrt{2\Omega_l}} e^{i\int^\eta \Omega_l d\eta} ,
\label{eq:WKBsol}
\end{equation}
with a further restriction,
\begin{equation}
\dot\chi_l = -i\Omega_l \left[\frac{\alpha_l}{\sqrt{2\Omega_l}} e^{-i\int^\eta \Omega_l d\eta} - \frac{\beta_l}{\sqrt{2\Omega_l}} e^{i\int^\eta \Omega_l d\eta}\right] ,\label{eq:WKBrestriction}
\end{equation}
where $\alpha_l$ and $\beta_l$ are Bogoliubov coefficients.

Putting Eqs.(\ref{eq:WKBsol}) and (\ref{eq:WKBrestriction}) in Eqs.(\ref{eq:genform}) and the condition  (\ref{eq:wronskian}), we get,
\begin{eqnarray}
\dot \alpha_l& = &\frac{1}{2} \left( \frac{\dot \Omega_l}{ \Omega_l} - i\frac{Q}{\Omega_l}\right)\beta_l\ e^{+2i\int\Omega_ld\eta} - i \frac{Q}{2 \Omega_l}\alpha_l  ,\nonumber\\
                                             \label {eq:alphabetadot}       \\
\dot \beta_l& = &\frac{1}{2} \left( \frac{\dot \Omega_l}{ \Omega_l} + i\frac{Q}{\Omega_l}\right)\alpha_l\ e^{-2i\int\Omega_ld\eta} + i \frac{Q}{2 \Omega_l}\beta_l ,\nonumber
\end{eqnarray}

and,\begin{equation}
|\alpha_l|^2 - |\beta_l|^2 = 1 .\label{eq:modsqrdiff}
\end{equation}

With the initial conditions,
\begin{equation}
 \alpha_l (\eta_0) = 1 \hspace{0.5cm} \mbox{and}\hspace{0.5cm} \beta_l(\eta_0) = 0, \label{eq:initial1}
\end{equation}
the number of particles created in mode $l$ ($l$ signifies both $k$ and $k_\sigma$) is given by,
\begin{equation}
N_l = \lim_{\eta\rightarrow \infty} |\beta_l|^2 .\label{eq:partno}
\end{equation}


Following Zel'dovich and Starobinsky \cite{zel'dovich}, Eqs.(\ref{eq:alphabetadot}) and (\ref{eq:modsqrdiff}), with the initial conditions (\ref{eq:initial1}), can be cast in the form,
\begin{eqnarray}
\frac{ds_l}{d\eta}& = & \frac{\dot \Omega_l}{2 \Omega_l}v_l + \frac{Q}{2 \Omega_l}r_l \nonumber, \\
\frac{dv_l}{d\eta}& = & \frac{\dot \Omega_l}{\Omega_l}(1 + 2s_l) - \left[\frac{Q}{\Omega_l} + 2 \Omega_l \right]r_l \label{eq:gennum},\\ 
\frac{dr_l}{d\eta}& = & \frac{Q}{\Omega_l}(1 + 2s_l) + \left[\frac{Q}{\Omega_l} + 2 \Omega_l \right]v_l ,\nonumber
\end{eqnarray}

with initial conditions,
\begin{equation}
s_l(\eta_0) = r_l(\eta_0) = v_l(\eta_0) = 0 ,\label{eq:ZSconds}
\end{equation}

where,
\begin{eqnarray}
s_l \equiv |\beta_l|^2,\nonumber
\end{eqnarray}
\begin{eqnarray}
v_l  \equiv 2 Re\left[ \alpha_l \beta_l^* e^{-2i\int^\eta \Omega_l d\eta}\right]\nonumber
\end{eqnarray}
and
\begin{eqnarray}
r_l \equiv 2i Im\left[ \alpha_l \beta_l^* e^{-2i\int^\eta \Omega_l d\eta}\right] .\nonumber
\end{eqnarray}

To get the number of particles created in mode $l$ one has to solve the first order differential system (\ref{eq:gennum}) with initial conditions (\ref{eq:ZSconds}) and determine $s_l$ when $\eta\rightarrow \infty $. These equations can be evolved numerically using standard, easily available codes, 
in cases where one is unable to find an analytic solution.

The number of created particles per unit volume, in ordinary space, in mode $k_\sigma$ is given at late times by
\begin{equation}
N_{k_\sigma} = \frac{1}{(2 \pi a)^3} \int {d^3k \ \vert \beta_l \vert ^2 } = \frac{1}{(2 \pi a)^3 } \int {4\pi k^2\vert \beta_l \vert ^2 }\ dk ,\label{eq:numden}
\end{equation}

The energy density is given by
\begin{equation}
\rho_{k_\sigma} = \frac{1}{(2 \pi a)^3 a} \int {d^3k \ k\vert \beta_l \vert ^2 } =  \frac{1}{(2 \pi a)^3 a} \int {4\pi k^3\vert \beta_l \vert ^2}\ dk .\label{eq:energyden}
\end{equation}


\section{The allowed values of $k_{\sigma}$}

We first note the fact that there are specific allowed values of $k_\sigma$ 
which depend on the nature of the warp factor and the boundary conditions. 
To obtain these allowed values of $k_\sigma$, we solved 
the Eq.(\ref{eq:separatingsigma}), which is an eigenvalue equation 
for $G(\sigma)$, for a typical functional form of $f(\sigma)$ (the RS solution). Other choices
of $f(\sigma)$ can also be studied in a similar way.

In Eq.(\ref{eq:separatingsigma}) let us take $G(\sigma) = G_1(\sigma)G_2(\sigma)$, the requirement of a vanishing coefficient in the term 
involving $G_1'(\sigma)$ leads to the choice,
\begin{equation}
G_2(\sigma) = const.\ e^{-f(\sigma)} ,\label{eq:G2}
\end{equation}
and, subsequently, we have
\begin{equation}
G_1''(\sigma) + \left[- \frac{f''}{2} - \frac{f'^2}{4} +  k_\sigma^2\ e^{-2f(\sigma)}\right]G_1(\sigma) = 0 .\label{eq:G1}
\end{equation}
Now we can treat Eq.(\ref{eq:G1}) as an eigenvalue equation for $G_1(\sigma)$ 
and investigate its solutions and the allowed values of $k_\sigma$. 

Following RS, we choose, $f(\sigma) = -\ b\ |\sigma|$. The 
Eq.(\ref{eq:G1}), for a two-brane model, now becomes
\begin{equation}
G_1''(\sigma) + \left[b\left(-\delta(\sigma) + \delta(\sigma - r_c\pi)\right) - \frac{b^2}{4} +  k_\sigma^2\ e^{2b|\sigma|}\right]G_1(\sigma) = 0 .\label{eq:G1case1}
\end{equation}
The complete solution of this differential equation is,
\begin{equation}
G_1(\sigma) = C\sqrt{\frac{2b\ e^{-b|\sigma|}}{k_\sigma}} \frac{cos\left[\frac{k_\sigma}{b}(\frac{e^{b|\sigma|}}{b} - 1) \right]}{cos(k_{\sigma}/b)} ,\hspace{0.5cm}\mbox{where $C$ is an arbitrary constant}.\label{eq:G1case1soln}
\end{equation}
 To crosscheck this result we use the following transformations, $z = sgn(\sigma)\ \frac{e^{b|\sigma|} - 1}{b}$ and $\tilde{G}(z) = G_1(\sigma)\ e^{\frac{b|\sigma|}{2}}$ (due to \cite{wbw3}) in Eq.(\ref{eq:G1case1}), which leads to,
\begin{equation}
\frac{d^2 \tilde{G}(z)}{dz^2} +  k_\sigma^2\ \tilde{G}(z) = 0 ,\label{eq:G1case1check}
\end{equation}
and confirms the previous result.
Now, for a two-brane model, the boundary conditions at $\sigma = 0$ and $\sigma = r_c \pi$ imply
\begin{equation}
  \frac{d\tilde{G}(z)}{dz} = -\frac{3}{2}\frac{b}{1 + b|z|}\tilde{G}(z) \vert_{z = 0 \hspace{0.3cm}\mbox{and\hspace{0.3cm}} z = \frac{e^{b r_c \pi} - 1}{b}}.\label{eq:boundaryconditions}
\end{equation}
Eq.(\ref{eq:G1case1check}) with the above boundary conditions completely 
determine the allowed values for $k_{\sigma}$ through the following 
transcendental equation
\begin{equation}
 \tan\left[\theta + \tan^{-1}\left(-\frac{2}{3}\theta e^{-b r_c \pi}\right)\right] = - \frac{2}{3}\theta \hspace{0.5cm}\mbox{where}\hspace{0.5cm} k_\sigma = \theta\ b\ e^{-b r_c \pi}.\label{eq:transcendental}
\end{equation}
We can approximate the above equation as,
\begin{equation}
 \tan\theta = - \frac{2}{3}\theta \hspace{0.3cm}\mbox{}\label{eq:transcendental1}
\end{equation}
for moderate values of $br_c$ (i.e. ignoring the second term in the
square brackets in the original transcendental equation). 

\begin{figure}[!ht]
\includegraphics[width = 4 in, height = 2.1 in]{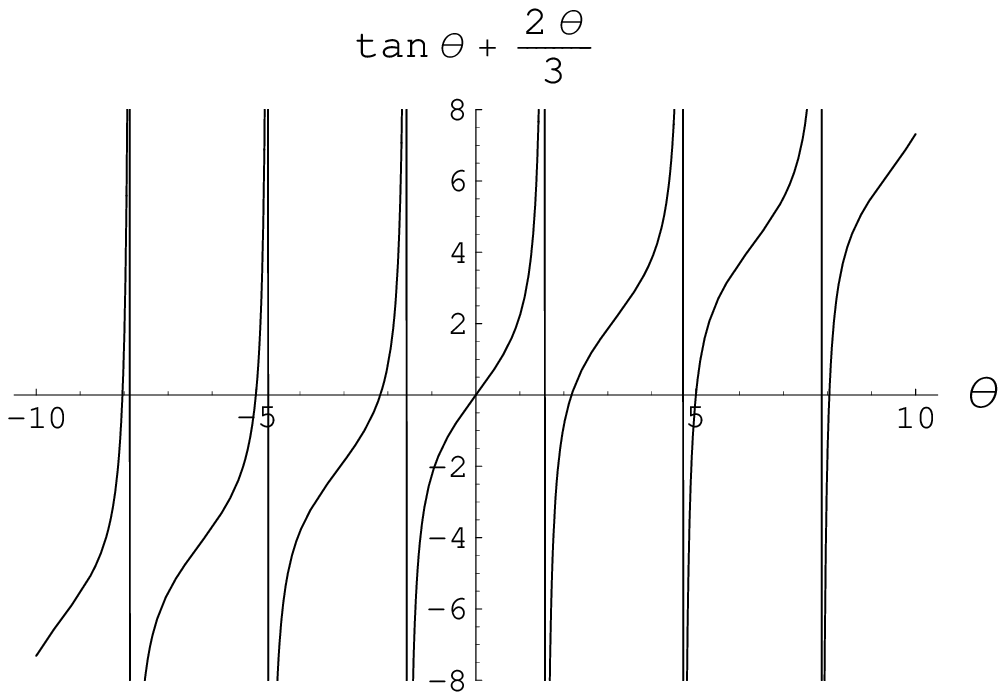}
\caption{} \label{fig:transcend}
\end{figure}

Solving graphically (see Fig. 1) we get $\theta = 0, \pm 2.17, \pm 5.01, \pm 8.04, \pm 11.13, \pm 14.24, \pm 17.36$ and so on. We consider $k_\sigma =\theta$
in TeV units (as is normally done in warped braneworld scenarios) 
by absorbing the additional factor in the definition of the energy unit. 

In the case of an infinite extra dimension there may or may not be any 
discrete values of the modes. Even, if there are some (bound states within 
the volcano-shaped potential), it is clear that for a decaying warp factor
of the RS type we would get a finite number of discrete values, and, beyond 
these values,
we will have a continuum. In our calculations henceforth, we shall consider
only the first three discrete values of $k_\sigma$ given above, for the
two--brane model.

\section{Particle creation in a 4d universe}

Let us first look at the known analysis of a conformally coupled, massive
scalar field in a four dimensional universe. 
In order to arrive at concrete results on quantities characterising 
particle creation we have to make a choice for the scale factor $a(\eta)$. 
This is chosen to be:
\begin{equation}
a^2 (\eta) = b_1^2 + b_2^2 \eta^2
\end{equation}
Note that the above choice gives a non-singular line element which has a bounce at $\eta = 0$ (Fig.\ref{fig:sf}). The
approach to the big--bang singularity can be modeled using the limit 
$b_1\rightarrow 0$. $b_2$ is known as the slowness parameter. 
In fact, for $b_1\rightarrow 0$ or $\eta\rightarrow\pm\infty$, the scale
factor approaches that of a radiative universe.

It is easy to show that in a four dimensional universe, the temporal part of
a massive, conformally 
scalar field in the background given by:
\begin{equation} 
ds^2 =\ a(\eta)^2(-\ d\eta^2 + dx^2 + dy^2 + dz^2)  ,\label{eq:4dmetric}
\end{equation}
satisfies the equation
\begin{equation}
\ddot \chi_l(\eta) + \Omega_l^2(\eta) \chi_l(\eta) = 0 ,\label{eq:eqn4d}
\end{equation}
 where
\begin{equation}
\Omega_l^2(\eta) = k^2 + (b_1^2 + b_2^2 \eta^2)m^2.\label{eq:omega4d}
\end{equation}
As shown in \cite{birrell}, one can construct an exact solution of the above 
equation in terms of parabolic cylinder functions. Using these solutions,
one can easily obtain the particle number density, 
\begin{equation}
|\beta_k|^2 = exp\left[-\pi \left(\frac{k^2}{mb_2} + \frac{mb_1^2}{b_2}\right)\right].\label{eq:beta4d}
\end{equation}

\begin{figure}[!ht]
\includegraphics[width = 4in]{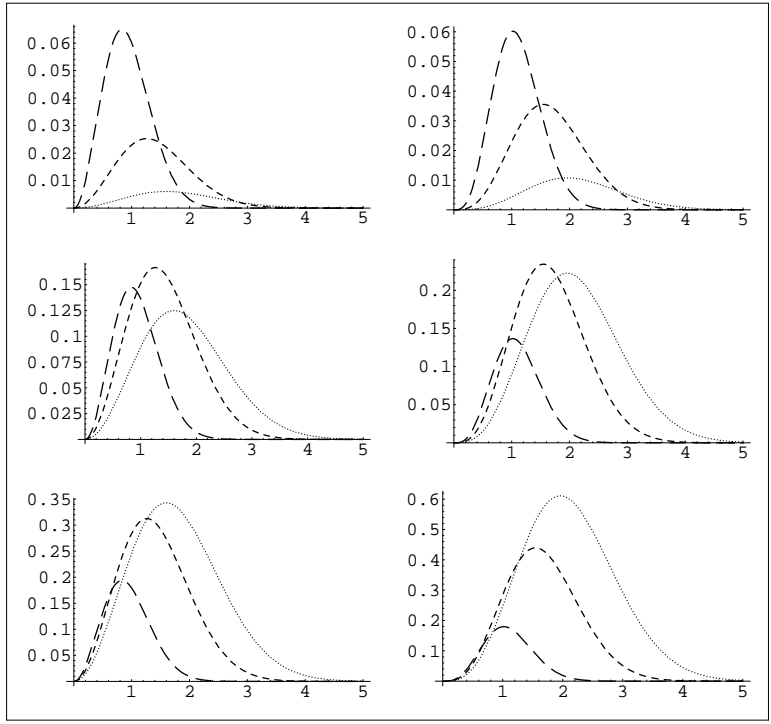}
\caption{Plots of $k^2 |\beta_k|^2$ and $k^3 |\beta_k|^2$ vs. $k$, respectively,
in first and second columns, for the analytic solution in a universe without 
any extra dimension, for 3 values of m such as $2.17$(long dashes), $5.01$(small dashes), $8.04$(dotted). Here $b_2 = 1$, and in the 1st row, $b_1 = \sqrt{0.2}$; in the 2nd row, $b_1 = \sqrt{0.08}$ and in the 3rd row, $b_1 = \sqrt{0.04}$.} \label{fig:pAn} 
\end{figure}
The main features of the particle number density ($k^2 |\beta_k|^2$) and 
 energy density ($k^3 |\beta_k|^2$), can be derived from Fig.\ref{fig:pAn}, are the following: 

\noindent (a) the $k$ value, where $k^2 |\beta_k|^2$ or $k^3 |\beta_k|^2$ peaks, 
is independent of the parameter $b_1$ (for same $m$) but increases with 
increasing $m$. 

\noindent (b) The heights of all the peaks increases with decreasing $b_1$.
For higher values of $b_1$, the lower modes dominate, whereas, when  we 
lower $b_1$ (ie. we get closer to a metric with a singularity at $\eta = 0$) 
the higher modes become more and more dominant. Also, it is known that massless 
particles ($m = 0$ modes) will not be created, as the background universe is  
conformally flat (check this using $m=0$ in the formula for
${\vert \beta_k \vert}^2$).  

\noindent (c) The nature of the spectrum of created particles can be
identified, following \cite{schafer}, with that of a non--relativistic thermal
gas of particles with momentum $\frac{k}{a}$ at a chemical potential
$-\frac{1}{2} \frac{m b_1^2}{a^2}$ and temperature $\frac{b_2}{2\pi a^2 k_B}$
($k_B$ is the Boltzmann constant).  

\section{Particle creation with a time-independent extra dimension}

Having looked at the four dimensional scenario and also introduced
the scale factor which we shall be working with throughout, we now move
on to the case of a time--independent extra dimension. 
Note that this is different from the usual Kaluza--Klein scenario 
because of warping and also because of the choice of the extra
dimensional space which could be finite or infinite.
A fair amount of work on such brane cosmological models (the 3-braneworlds
are now the so--called FRW branes) has been carried out since the
inception of the braneworld idea \cite{branecosmo}.

\subsection{Approximate analytic solution}

The analysis of a massless scalar field conformally coupled with a 5d metric 
(\ref{eq:metric}) is now carried out with the choice 
$\phi (\eta) = b_1$ (a constant 
equal to the minimum value of the scale factor $a(\eta)$). 
The Eq.(\ref{eq:genchieqneta}) takes the form:
\begin{equation}
\ddot \chi_l(\eta) + \left[ \left(k^2 + \frac{b_1^2 + b_2^2 \eta^2}{b_1^2}k_\sigma^2\right) + \frac{b_1^2b_2^2}{8(b_1^2 + b_2^2\eta^2)}\right]\chi_l(\eta) = 0 ,\label{eq:eqn5dcon}
\end{equation}
We have not succeeded in finding an analytic solution of the above equation. 
But, Eq.(\ref{eq:eqn5dcon}) and Eq.(\ref{eq:eqn4d}), are similar (modulo
some redefinitions), 
if we consider that the  particle concept has a meaning only 
at $\eta\rightarrow\pm\infty$. In this case, the second term in the
square brackets in Eq. (5.1) can be ignored and an approximate solution 
can thus be found which is the same as that obtained in the scenario of a
four dimensional universe. This leads to
\begin{equation}
|\beta_k|^2 = exp\left[-\pi \left(\frac{k^2b_1}{k_\sigma b_2} + \frac{k_\sigma b_1}{b_2}\right)\right].\label{eq:beta5dcon}
\end{equation}
Hence, the particle number density becomes dependent on the size of the extra 
dimension. This approximation implies the equivalence of $m$ and $\frac{k_\sigma}{b_1}$, i.e. one may imagine a massive field in 4d as a projection of a 
massless scalar field residing in the 5d bulk. The variations of total particle number density and total energy density w.r.t $k$ for different parameter 
dependences are plotted in the left box of Fig.\ref{fig:PN&peqAn}. 
The following features can be noted:

\noindent (a) The $k$ value where $k^2 |\beta_k|^2$ or $k^3 |\beta_k|^2$ peaks 
increases with decreasing $b_1$ for same $k_\sigma$ and it also increases with 
increasing $k_\sigma$. 

\noindent (b) The heights of all the peaks increases with decreasing $b_1$. 

\noindent (c) The higher modes dominate for lower $b_1$ values.

\noindent (d)  There will be 
no $k_\sigma = 0$ modes created (unlike the 4D case).
Recall that the conformal invariance is not broken in the field equation since  
we have taken the $Q(\eta)$ factor to be negligible.

\subsection{Numerical solution}

The case discussed in the previous subsection is now re-analysed using the
Zel'dovich-Starobinsky equations given in the previous section (taking 
into account the extra factor $Q(\eta)$) and the corresponding variations 
are plotted in the right box of Fig.\ref{fig:PN&peqAn} in order to 
check how good the approximate analytic solution is.

\begin{figure}[!ht]
\includegraphics[width=3.2in]{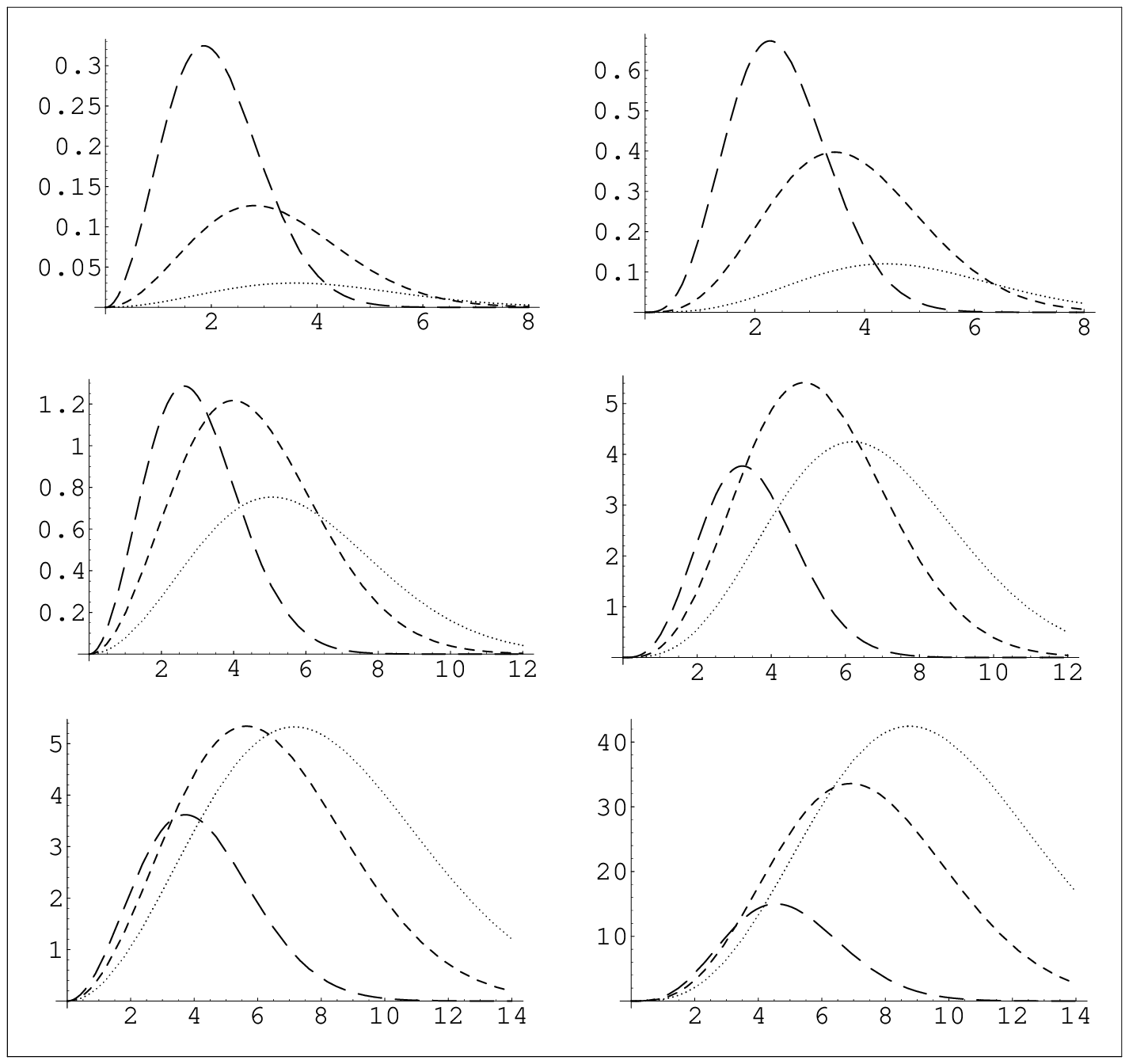}
\includegraphics[width=3.2in]{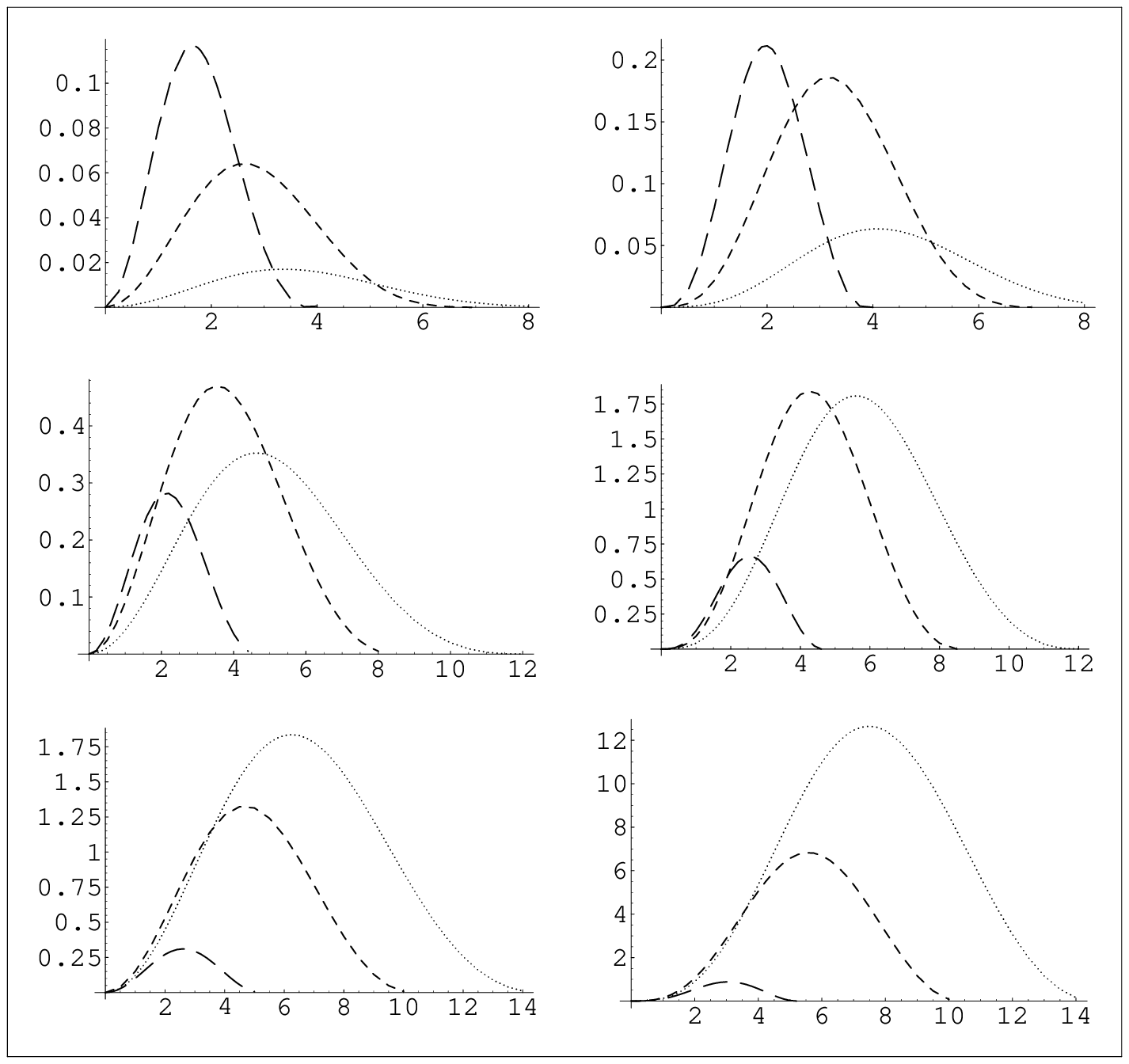}
\caption{Left box shows plots of the analytic solution and the right box depicts numerically evaluated $k^2 |\beta_k|^2$ and $k^3 |\beta_k|^2$ vs. $k$ in two columns in a universe with a static extra dimension (with $b_2 = 1$), for 3 values of $k_\sigma$, such as $2.17$(long dashes), $5.01$(small dashes), $8.03$(dotted), where in 1st row, $b_1 = 0.2$; in 2nd row, $b_1 = 0.1$ and in 3rd row, $b_1 = 0.05$. } \label{fig:PN&peqAn} 
\end{figure}

The differences that arise between the results in this and those quoted
in the previous subsection are: 

\noindent (a) For same $k_\sigma$ and $b_1$, peaks occur at different $k$ 
values.

\noindent (b) The presence of a static extra dimension decreases the amplitudes 
of total particle number density and energy density (this is a signature of the $Q (\eta)$ term). Here, it is obvious that zero mode particles will be 
created, but this phenomenon is addressed later.

\noindent (c) The spectrum is no longer that of a thermal gas of non--relativistic particles. This is evident from the figure above where we show the
$k^2{\vert \beta_k \vert}^2$ and $k^3{\vert \beta_k \vert}^2$ for the 
approximate and the numerical analysis
carried out in this section. Although it might seem from the plots of the number
and energy densities that the spectrum is thermal, we note that this is
not the case. The obvious reason behind this is hidden in the $Q(\eta)$
factor which breaks the conformal symmetry and thereby leads to a non--thermal
spectrum. One might take this as a signature of the presence of the
extra dimension. Our results with a time--dependent extra dimension
which we shall turn to now, also exhibit the same fact. 

\section{Particle creation with a time dependent extra dimension}

\begin{figure}[!ht]
\includegraphics[width = 5in]{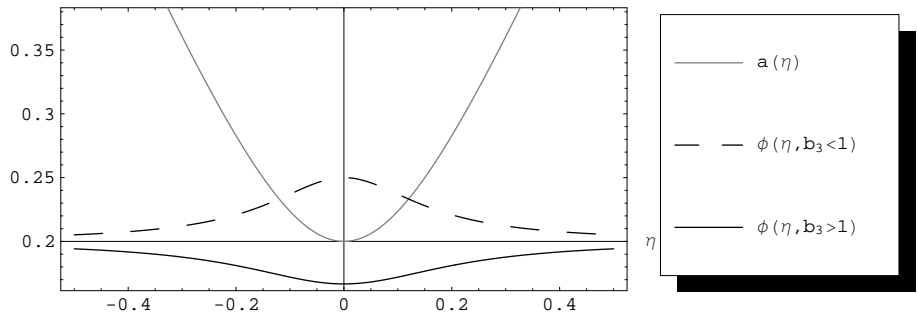}
\caption{The above plot shows the variation of $a(\eta)$ and $\phi(\eta)$ (for two different choices of $b_3$; we have taken $b_1 = 0.2$, $b_2 = 1$, $b_3 = 1.2$ and $0.8$ )} \label{fig:sf} 
\end{figure}

What happens when 
$\phi(\eta)$ is time-dependent? Such extra dimensions have not
really been studied much in the context of braneworld models. The
unwarped scenario with time dependent extra dimension has been
analysed in \cite{time} with the motivation of obtaining an
accelerating scale factor. Exact
solutions of the Einstein equations with a non-constant $\phi(\eta)$ and
with physically motivated matter
sources are rare and difficult to find, though some examples and their
cosmological implications have been discussed in \cite{freese}. 
  However, in the Kaluza--Klein
context time dependent extra dimensions which decay in time along with
the expansion of the universe have been dealt with in detail (see papers
on Kaluza--Klein cosmology in \cite{mkkt}). 

Here, we assume
the same form (as in the previous sections) for the scale factor $a(\eta)$. The $\phi(\eta)$ is chosen
to be a function of $\eta$ which approaches the constant value $b_1$ 
at the asymptotic infinities ($\pm \infty$). The form we choose is:
 
 \begin{equation}
\phi^2(\eta) = \frac{b_1^2 + b_2^2\eta^2}{b_3^2 + b_2^2 \eta^2/b_1^2}.\label{eq:phieta}
\end{equation}
 Fig.\ref{fig:sf} shows the variations of the scale factor for typical parameter values. $\phi(\eta)$ in both cases ($b_3>1$ and $b_3<1$) goes over to $b_1$ as $\eta\rightarrow\infty$ (which was the size of the extra dimension considered
in the previous section).

\begin{figure}[!ht]
\includegraphics[width = 4in]{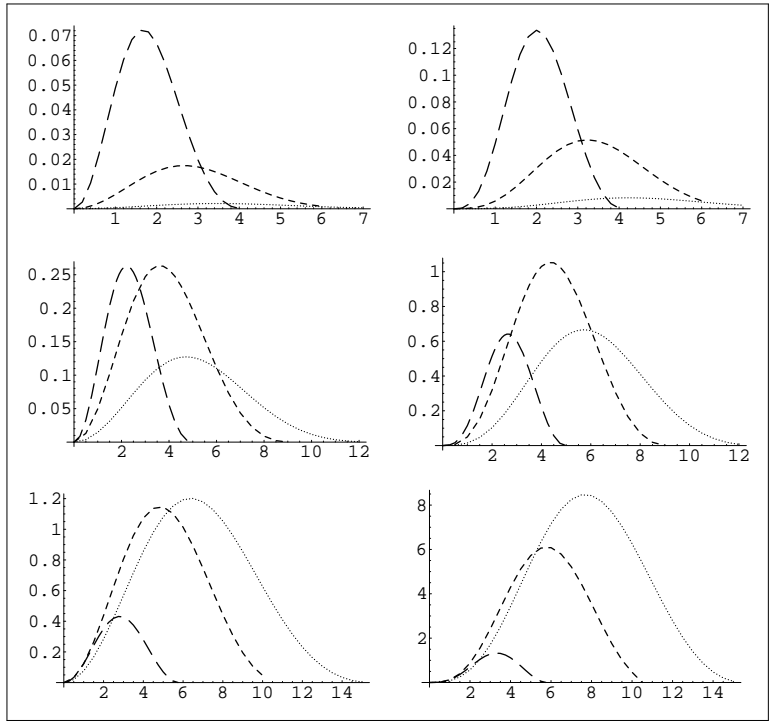}
\caption{Plots of numerically evaluated $k^2 |\beta_k|^2$ and $k^3 |\beta_k|^2$ vs. $k$ in two columns, in a universe with a dynamic extra dimension (with $b_3 = 1.2$ and $b_2 = 1$) for 3 values of $k_\sigma$ such as $2.17$(long dashes), $5.01$(small dashes), $8.04$(dotted), where in 1st row, $b_1 = 0.2$; in 2nd row, $b_1 = 0.1$ and in 3rd row, $b_1 = 0.05$. } \label{fig:PNvx} 
\end{figure} 
The particle number density and energy density are evaluated numerically
(as before) in 
the presence of the above form of a time-varying extra dimension, 
for a typical value of $b_3$ $(>1)$ and plotted in Fig.\ref{fig:PNvx}.
In this case, the size of the time--varying extra dimension always 
remains 
less than the minimum size of ordinary space. Apart from the features pointed 
out earlier, we can see that for $b_3>1$, the contribution of the higher modes 
diminishes very quickly 
as compared to the previous cases. The reverse is seen to happen for $b_3<1$,
i.e. the higher modes tend to become more and more dominant with decreasing values of $b_3$. Both these results are however dependent on the choice of
$b_1$ which is taken to be large enough (away from the approach to
a singular $a(\eta)$ ($b_1\rightarrow 0$)). But, for a fixed $b_1$ 
(say $b_1=0.2$) the marked differences in the number and energy densities 
for $b_3>1$ and $b_3<1$ are clearly visible from the next plots. 
 
\subsection{Comparison}

In order to figure out the similarities and differences arising out 
of a time-dependent and independent extra dimension, we now make a
comparison of the results of the previous subsections. The differences are 
mentioned pointwise below.

\begin{figure}[!ht]
\includegraphics[width = 4in]{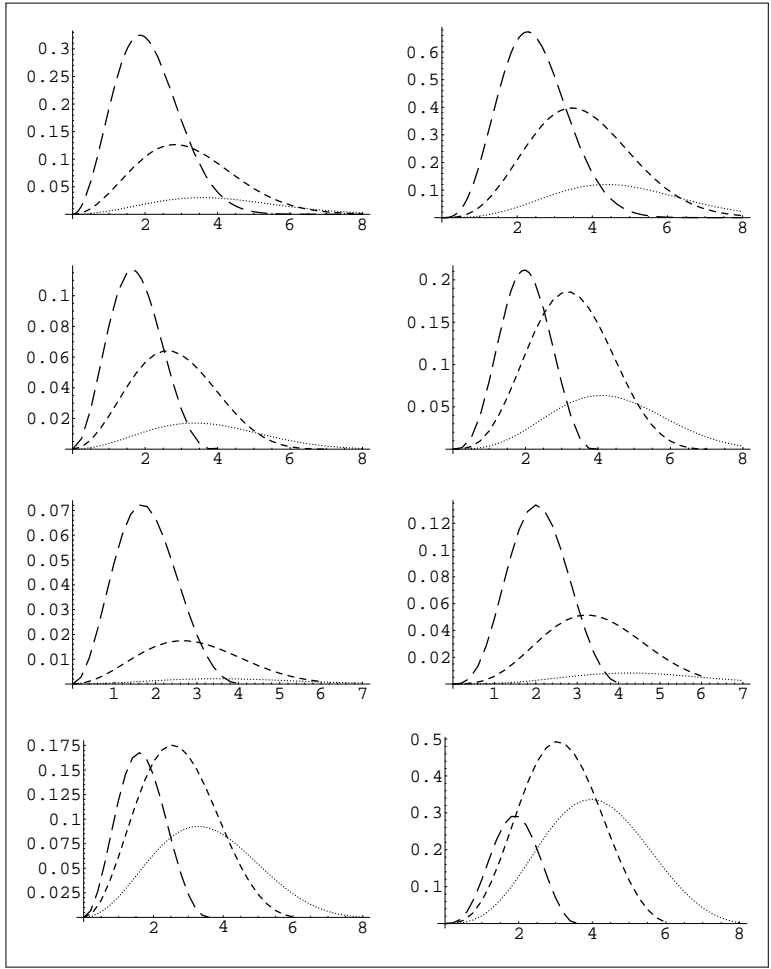}
\caption{Comparison of plots of $k^2 |\beta_k|^2$ and $k^3 |\beta_k|^2$ vs. $k$ in two columns with, $b_1 = 0.2$ and $b_2 = 1$, in a universe - 1st row: without extra dimension and $m = \frac{ k_\sigma}{b_1}$; 2nd row: with static extra dimension; 3rd row: with dynamic extra dimension and $b_3 = 1.2$; 4th row: with dynamic extra dimension and $b_3 = 0.8$; for 3 values of $k_\sigma$ such as $2.17$(long dashes), $5.01$(small dashes), $8.04$(dotted). } \label{fig:comp} 
\end{figure}
In Fig.\ref{fig:comp}, as the caption suggests, we have placed the
salient features of the integrands required to compute the total particle 
number density and the total energy density in the four different situations 
(universe without extra dimension, with static extra dimension, 
and with a dynamic extra dimension of two different kinds). We use the
 same value of the parameters $b_1$ (determining the minimum size of the 
ordinary scale factor) and $b_2$ (the slowness parameter) and the same 
$k_\sigma$ values (which are the first three non-zero solutions found in 
Section III) in consecutive rows.  

\noindent (a) The presence of a static extra dimension decreases the amplitudes 
for each mode, but the higher modes become relatively more dominating 
(i.e. relative contribution of higher modes w.r.t lower modes increases) in 
comparison with the case of a 4d universe.

\noindent (b) We have two different kinds of time-varying extra dimension 
(Fig.\ref{fig:sf}). If the size of the extra dimension always remains less 
than the value equal to the static case and asymptotically reaches the same 
value (see $\phi(\eta, b_3>1)$) in Fig.\ref{fig:sf}), {\em contributions of higher modes 
are significantly diminished and heights of the peaks are also decreased 
compared to the other cases}. However, for the extra dimension with 
$\phi(\eta, b_3<1)$, an exactly opposite effect is seen to happen.

\noindent (c) 
In the last three rows, the $k$-values where  
$k^2 |\beta_k|^2$ or $k^3 |\beta_k|^2$ peaks are the same and less than that 
in the 4d case.

\noindent (d) Previously, in models where the extra dimension was taken to be 
hidden and a decreasing function of time \cite{garriga}, any nonzero 
$k_\sigma$ mode, if excited, was found to soon dominate over the redshifted 
$k$-modes because of ever-increasing blueshift (their physical frequency 
asymptotically reaches a value equal to $k_\sigma/b_1$, but in the braneworld 
scenario $b_1$ can be much larger compared to those Planck-sized 
extra dimensions) resulting in a breakdown of the background cosmological 
model due to the possibility of a large back reaction. 
This problem may be resolved here for a dynamic extra dimension with $b_3>1$, 
where the zero-mode is more dominant and higher mode contributions are quickly 
suppressed i.e production of [$k, k_\sigma\neq0$] particles can be controlled
using suitable values of $b_3$.

\section{Non--thermal nature of the spectrum}

In the cases of the massive, conformally coupled scalar field in four
dimensions and the analytic treatment of the time independent extra dimension
(assuming $Q(\eta)=0$) the spectrum of created particles can be
identified with that of a non--relativistic thermal gas of particles.
However, including the $Q(\eta)$ factor we find that the spectrum
deviates from the thermal nature--we might call it nearly thermal though. 
The same features appear when we
consider time--dependent extra dimensions. In this section, we outline
these features by attempting to fit (by least-square fitting method) the 
numerical data using thermal as well as non--thermal profiles characterised
by a set of parameters. The results are shown below.\\
In the case of a static extra dimension we attempt to fit total particle 
number densities (top-left plot in right box of Fig.\ref{fig:PN&peqAn}) 
with the following expressions for thermal and nonthermal profiles 
of the particle number densities,
\begin{equation}
|\beta_k|^2 = A exp\left[-\pi \left(B\frac{k^2b_1}{k_\sigma b_2} + \frac{k_\sigma b_1}{b_2}\right)\right].\label{eq:thermalfit}
\end{equation}
\begin{equation}
|\beta_k|^2 = A exp\left[-\pi \left(B\frac{k^nb_1}{k_\sigma b_2} + \frac{k_\sigma b_1}{b_2}\right)\right].\label{eq:nonthermalfit}
\end{equation}
The first one is essentially the same thermal profile 
as Eq.(\ref{eq:beta5dcon}) with two new (fitting) parameters ($A$ and $B$) 
introduced. The second one is a nonthermal profile with another parameter 
($n$) introduced to take care of the non-thermality. The best-fit plots are 
shown in Fig.\ref{fig:thermalfit} and Fig.\ref{fig:nonthermalfit}. \\
(a) Deviations of the datapoints from thermal fits are very prominent for 
lower values of $k_\sigma$, whereas for higher and higher values of 
$k_\sigma$ data points seem to converge on the thermal fits. These features
are clearer when we use a non-thermal fit which seems to be a better fit 
in all the three cases, but the extent of non-thermality decreases with 
increasing $k_\sigma$ (i.e., value of $n$ converges towards $n=2$), this is due to the fact that the factor $Q/\Omega$ contributes lesser and lesser for larger and larger $k_\sigma$.\\
(b) The fall--off of $k^2|\beta_k|^2$, for large k, are much faster than the thermal fits. Again this is confirmed through non-thermal fits where $n$ is found to be greater than $2$.\\
(c) To figure out a single possible fitting formula for the non-thermal 
profile with a unique set of parameters 
one needs to do a rigorous statistical analysis. 
The immediately apparent features of the parameters 
in our fitting functions are the following. With increasing $k_\sigma$ we
find: $A$ increasing, $B$ showing an oscillatory behaviour and $n$  
converging towards $2$. These patterns can be useful in determining
the nature of non-thermality--an aspect which seems to be crucial
in distinguishing between the presence and absence of extra dimensions. 
\begin{figure}[!ht]
\includegraphics[width=3in,height=4.5in,angle=-90]{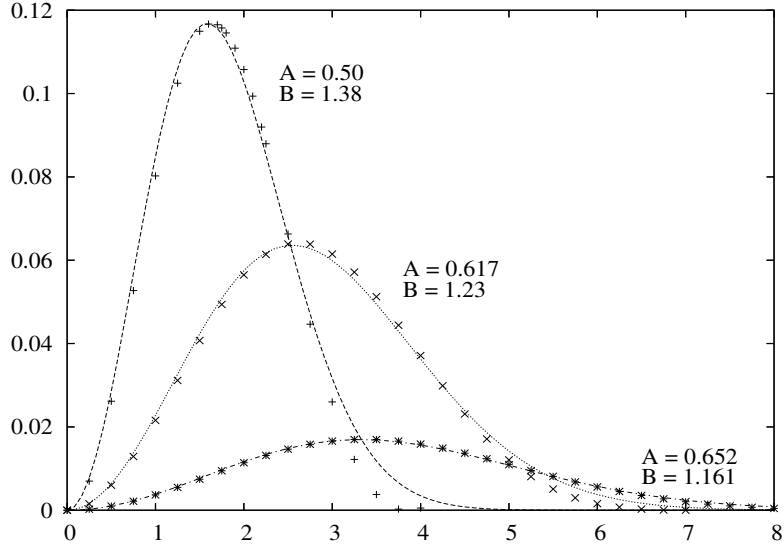}
\caption{Best-fit plots using a thermal profile for datapoints of $k^2|\beta_k|^2$ vs. $k$ (top-left plot in right box of Fig.\ref{fig:PN&peqAn} where $b_1 = 0.2$$, b_2 = 1$, for 3 values of $k_\sigma$, such as $2.17$(upper curve), $5.01$(middle curve) and $8.03$(lower curve)), and the corresponding best-fit parameter values.} \label{fig:thermalfit}
\end{figure}
\begin{figure}[!ht]
\includegraphics[width=3in,height=4.5in,angle=-90]{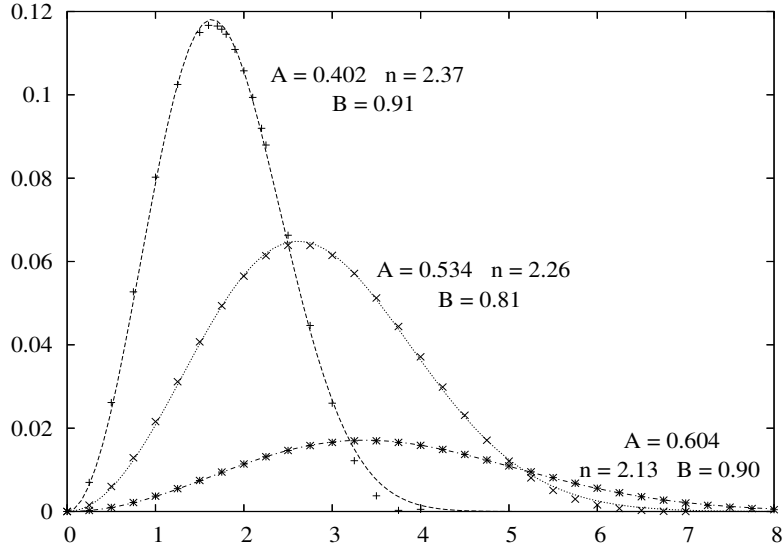}
\caption{Best-fit plots for a nonthermal profile for datapoints of $k^2|\beta_k|^2$ vs. $k$ (top-left plot in right box of Fig.\ref{fig:PN&peqAn} where $b_1 = 0.2$$, b_2 = 1$, for 3 values of $k_\sigma$, such as $2.17$(upper curve), $5.01$(middle curve) and $8.04$(lower curve)), and the corresponding best-fitting 
parameter values.} \label{fig:nonthermalfit}
\end{figure}

\section{The zero mode}

In \cite{garriga}, it is pointed out that, in a radiative universe 
$k_\sigma = 0$ modes or the massless particles in the four dimensional picture 
will not be created if we have a static extra dimension because 
this situation is equivalent to the problem of massless particle creation in a 
conformally flat four dimensional cosmology. This accidental recovery of 
conformal invariance happens only because of the particular choice of 
$a(\eta)$ ($\sim \eta $, in radiative universe) with a static extra dimension. 
Even in the case of a time-varying extra dimension conformal invariance can be 
imposed by making $Q(\eta) = 0$ (the trivial case is, when $a(\eta) = \phi(\eta)$, which is not of any interest). Otherwise, the conformal invariance is 
always broken for a metric of type Eq.(\ref{eq:metric}) and the zero mode 
particles will indeed be produced with cosmological evolution. 
In Fig.\ref{fig:zeromode}, features of zero mode particle creation in 
presence of a static extra dimension (continuous line) is compared with the
other two situations with different kinds of extra dimensions ($b_3>1$: smaller dashes and $b_3<1$: longer dashes).

\begin{figure}[!ht]
\includegraphics[width = 6 in]{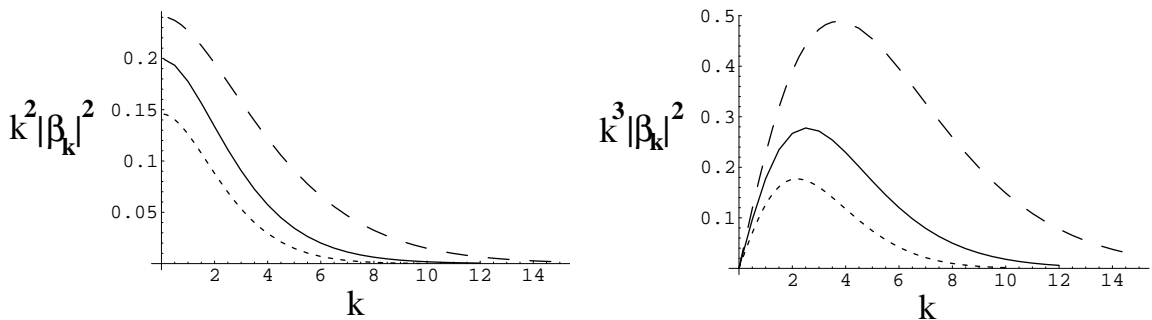}
\caption{Features of zero mode particle creation in presence of - static extra dimension (continuous line), dynamic extra dimension with $b_3 = 1.2$ (smaller dashes) and dynamic extra dimension with $b_3 = 0.8$ (longer dashes); when $b_1 = 0.2$ and $b_2 = 1$.} \label{fig:zeromode} 
\end{figure} 
The contribution of zero mode particles, in the static case, lies between 
those of the two dynamic cases (this feature is similar for the other modes). 
For these particular choices of parameters, the zero modes are the most 
dominating modes. But the plots of the integrand in the
total particle number density, asymptotically reaches a particular value 
as $k \rightarrow 0$ (for the zero modes), 
which is not the case for $k_\sigma \neq 0$ modes. 
Moreover, nothing similar to these modes (as the $m = 0$ modes) 
exist in a four dimensional universe (without any extra dimension), 
since such a geometry is conformally flat.

\section{Conclusions and comments}

In this article, we have investigated the possible effects of the presence 
of a warped extra dimension (static or dynamic) on particle production in a 
4d universe. 
From the numerical analysis represented via the figures, 
it is easy to point out the characteristic features.

$\bullet$ The warp factor plays its part through the values of $k_\sigma$
which are the allowed modes arising because of the warped extra dimension. 
$k_\sigma$ will have discrete values in the case of a two brane model, 
otherwise it will be a continuum. 

$\bullet$ The location and heights of the peaks in the plots for the
particle number density and energy density (as a function of $k$) 
decreases in the presence of an extra dimension. This is a feature of 
the breaking of conformal invariance and is manifest
 through the presence of the $Q(\eta)$ factor. 

$\bullet$ For the scenario with a time-varying extra dimension we have looked
at two types distinguished through the parameter $b_3$ in the expression for
$\phi (\eta)$.  We note an increase in the heights of the peaks 
for $\phi(\eta, b_3<1)$ and a decrease for $\phi(\eta, b_3>1)$). 
To understand further general features of the breaking of conformal invariance 
in the context of particle creation one needs to see what happens for 
different kinds of models (i.e choices of $a(\eta)$ and $\phi(\eta)$ 
which yield different forms for $Q(\eta)$).

$\bullet$ It may be mentioned that though these features look to be case 
specific
and model dependent, it is worth noting that we did not 
choose $\phi(\eta)$ arbitrarily. Rather, as mentioned before,
it is chosen to be equal to $b_1$ (avoiding the introduction of
another new parameter) or asymptotically reaching the same value
(in the dynamic cases). Moreover, this value is the minimum value of
 $a(\eta)$ during its evolution. In effect we are assuming some correlation 
between $a(\eta)$ and $\phi(\eta)$, i.e. there may be a single mechanism which 
drives both the scale factors and therefore their ought to be a connection
between them. This aspect can become clearer if we try to find out such 
solutions of the Einstein equations with a driving source in the bulk.

$\bullet$ We have also found a mechanism (for $\phi(\eta, b_3>1)$) to have 
control over contributions of higher non-zero $k_\sigma$ modes to the 
energy density of the universe. We can suppress the production of those 
higher modes at our will by choosing higher and higher value of $b_3$.  
The back reaction thus remains negligible and the classical cosmological model 
which is our background for all these studies does not
break down.

$\bullet$ To get an idea about the thermal/non--thermal nature of the
spectrum of created particles we have tried to fit our numerical data
with a thermal profile with a new set of parameters. Our analysis
reveals that the spectrum could be assumed to be nearly thermal
though marked deviations seem to appear for larger values of $k$.

$\bullet$ As a next step, one should study particle creation in the 
context of more realistic cosmological models and also 
further investigate the possibility
of creation of other types of particles (like fermions, or spin one particles
etc.) on the brane. Also, studies on particle production along the brane
could provide insights into the recently proposed idea of
braneworld isotropisation \cite{padilla}. 

We admit, in conclusion, that our results are based on a toy
scenario and can only be thought of as a build-up towards the study
of more relevant and realistic situations in future.
However, we do believe that some of the features observed (eg. possible
avoidance of significant back--reaction effects and non-thermal nature
of the spectral profile) here will
indeed be carried over in generically similar situations and can
provide pointers towards a better understanding of quantum fields in the
presence of a warped extra dimension.

\section*{Acknowledgements}
The authors thank G. Niz, A. Padilla and F. Urban for making them aware
of related works and also for their useful comments. SG thanks IIT Kharagpur,
India for providing financial support and the Centre for
Theoretical Studies, IIT Kharagpur, India for allowing him to use 
its research facilities.


\begin{thebibliography}{99}

\bibitem{kk} Th. Kaluza, Sitzunober. Preuss. Akad. Wiss. Berlin, p.966 (1921); 
O. Klein, Z. Phys. {\bf 37} 895 (1926).

\bibitem{wbw1} G. Gogberashvili, Int. J. of Mod. Phys. D {\bf 11}, 1635(2002).
\bibitem{wbw2} L. Randall and R. Sundrum, Phys. Rev. lett. {\bf 83} 3370 (1999).
\bibitem{wbw3} L. Randall and R. Sundrum, Phys. Rev. lett. {\bf 83} 4690 (1999).
\bibitem{wbw4} N. Arkani-Hamed, {\em et al.} Phys. Rev. lett. {\bf 84} 586 
(2000). 

\bibitem{loc} B. Bajc and G. Gabadadze, Phys. Lett. {\bf B 474}, 282 (2000);
S. Randjbar-Daemi and M. Shaposhnikov, Phys. Lett. {\bf B 492}, 361 (2000);
S. L. Dubovsky, V. A. Rubakov and P. G. Tinyakov, Phys. Rev. {\bf D 62},
105011 (2000);
Y. Grossman and N. Neubert, Phys. Lett. {\bf B 474} 361 (2000);
C. Ringeval, P. Peter, J. P. Uzan, Phys. Rev. {\bf D 65}, 044416 (2002);
S. Ichinose, Phys. Rev. {\bf D 66}, 104015 (2002);
R. Koley and S. Kar, Class. Quantum Grav. {\bf 22}, 753 (2005);

\bibitem{garriga} J. Garriga and E. Verdaguer, Phys. Rev. D {\bf 39} 1072 (1989).

\bibitem{nojiri} S. Nojiri and S. D. Odintsov, JCAP {\bf 06} 004 (2003).

\bibitem{makharko} M. K. Mak and T. Harko, Class. Quantum Grav. {\bf 16} 4085 (1999).

\bibitem{huang} W. H. Huang, Phys. Lett. A{\bf 140} 280 (1989).

 \bibitem{saharian} A. A. Saharian  Phys. Rev. D {\bf 73} 44012 (2006).

\bibitem{urban} C. Bambi and F. R. Urban,  Phys. Rev. Lett.  
{\bf 99} 191302 (2007).


\bibitem{birrell} N. D. Birrell and P. C. W. Davies, {\em Quantum fields in curved space} (Cambridge University press, Cambridge, 1982) and references therein.

\bibitem{zel'dovich} Ya. B. Zel'dovich and A. A. Starobinsky, Zh. Eksp. Teor. Fiz. {\bf 61}, 617, (1981) [Sov. Phys. JETP {\bf 34}, 1159 (1972)].

\bibitem{schafer} J. Audretch and G. Sch\"{a}fer, Phys. Lett. A {\bf 66} 459 (1978).

\bibitem{branecosmo} 
P. Binetruy, C. Deffayet, D. Langlois,
Nucl.Phys. {\bf B565} 269 (2000); 
P. Binetruy, C. Deffayet, U. Ellwanger, D. Langlois, Phys.Lett. {\bf B477} 
285 (2000); N. Kaloper, Phys.Rev. {\bf D60} 123506 (1999); 
P. Bowcock, C. Charmousis, R. Gregory, Class.Quant.Grav. {\bf 17}, 
4745 (2000); P. Brax and C. van de Bruck, Class.Quant.Grav. {\bf 20} R201 
(2003) 

\bibitem{time}
Je-An Gu, W-Y. P. Hwang, Phys.Rev. {\bf D66} 024003 (2002), K. Freese and M.Lewis, Phys. Letts. {\bf B540}, 1 (2002);
J. Cline and J. Vinet, Phys. Rev. {\bf D68} 025015 (2003)

\bibitem{freese} D. J. H. Chung and K. Freese, Phys.Rev. {\bf D 61} 023511 (2000); A. Wong, R-G Cai and N. O. Santos,  Nucl.Phys. {\bf B797}, 395 (2008).

\bibitem{mkkt} T. Appelquist, A. Chodos and P. G. O. Freund, {\em Modern
Kaluza--Klein theories} (Addison Wesley Publishing Company, 1987) 


\bibitem{padilla} G. Niz, A. Padilla and H. K. Kunduri, arXiv:0801.3462 
\end{thebibliography}
\end{document}